\documentclass{article}
\usepackage[totalwidth=440pt, totalheight=560pt]{geometry}
\usepackage{amssymb,amsmath}
\usepackage{amsfonts}
\usepackage{epsfig}
\usepackage{graphicx}
\usepackage{natbib}

\numberwithin{equation}{section}

\def\d{{\rm d}}
\newcommand{\p}{\partial}
\def\tbs{\textbullet\hspace{3mm}}
\newcommand{\bv}{{\mbox{\boldmath $v$}}}
\newcommand{\br}{{\mbox{\boldmath $r$}}}
\newcommand{\case}{\textstyle\frac}

\begin{document}

\title{Notes on the stability threshold \\for radially anisotropic polytropes}
\author{\small E.~V.~Polyachenko,$^1$\thanks{E-mail: epolyach@inasan.ru} V. L. Polyachenko,$^1$
        I. G. Shukhman,$^2$\thanks{E-mail: shukhman@iszf.irk.ru} \\[2mm]
       $^1${\it\small Institute of Astronomy, Russian Academy of Sciences,} \\
           {\it\small 48 Pyatnitskya St., Moscow 119017, Russia} \\[2mm]
       $^2${\it\small Institute of Solar-Terrestrial Physics, Russian Academy of Sciences,} \\
           {\it\small Siberian Branch, P.O. Box 291, Irkutsk 664033, Russia}}



\maketitle

\begin{abstract}
We discuss some contradictions found in the literature concerning the
problem of stability of collisionless spherical stellar systems which
are the simplest anisotropic generalization of the well-known
polytrope models. Their distribution function  $F(E,L)$ is a product of
power-low functions of the energy $E$ and the angular momentum $L$, i.e.
$F\propto L^{-s}(-E)^q$. On the one hand, calculation of the growth
rates in the framework of linear stability theory and N-body
simulations show that these systems become stable when the parameter
$s$ characterizing the velocity anisotropy of the stellar distribution
is lower than some finite threshold  value, $s<s_\textrm{crit}$. On the
other hand  Palmer \& Papaloizou (1987) showed that the instability
remained up to the isotropic limit $s=0$.

Using our method of determining the eigenmodes for stellar
systems, we show that the growth rates in weakly
radially-anisotropic systems are indeed positive, but decrease
exponentially as the parameter $s$ approaches zero, i.e. $
\gamma\propto \exp(-s_{\ast}/s)$. In fact, for the systems with
finite lifetime this means stability.

\end{abstract}

\vspace{10mm}

\section{Introduction}

Stability properties of stellar spherical clusters determine a set of dynamically
allowable equilibrium configurations. Presence of an instability can,
for example, lead to ellipsoidal deformation.

For a long time, one believed that any stellar spherical clusters,
except for the pathological models, were stable. This belief
appeared after the classical works by Antonov (1960, 1962) devoted to
isotropic systems, and was reassured in the following papers concerning
some particular anisotropic systems (e.g., Mikhailovsky et al. 1970,
Doremus et al. 1971). So, it needed some time to realize the very
possibility of instability of spheres, starting from Polyachenko \&
Shukhman (1972) until Merritt \& Aguilar (1985), Barnes et al.
(1986) and May \& Binney (1986) have  made it widely known.

The radial orbit instability is suppressed for sufficiently rounded
orbits, or when the kinetic energy stored in transverse directions
$T_\perp$ becomes sufficiently high. Polyachenko \& Shukhman (1981)
proposed a global anisotropy parameter as a ratio of the radial to
transverse kinetic energy of the system, $\xi \equiv 2T_r/T_\perp$.
For the Idlis model (Idlis 1956), they found the stability boundary
$\xi = 1.59$. Later, Fridman \& Polyachenko (1984), using this and two
other families of models, proposed a hypothesis that the global
anisotropy parameter can give a general stability criterion for
anisotropic systems: the system is stable if $\xi < 1.7 \pm
0.25$.\footnote{Results of stability analysis of one of three families
of models described in Fridman \& Polyachenko (1984) were reconsidered
later in Polyachenko (1987) report. Reconsidered stability boundary
$2T_r/T_{\perp}$ fell between 2.05 and 2.10, instead of the
previous boundary, $2T_r/T_{\perp}=1.62$,  i. e. the systems proved to
be more stable than it was supposed before.   The same boundary for
this family was obtained later by Dejonghe and Merritt (1988) with the
help of N-body simulations. A new (corrected) boundary is slightly out
of the range  $2T_r/T_{\perp}=1.7\pm 0.25$ suggested by Fridman \&
Polyachenko (1984).}

Following studies of spherical models by means of the linear
stability analysis (Saha 1991, Weinberg 1991, Bertin et al. 1994)
and N-body simulations (Merritt \& Aguilar 1985, Barnes et al.
1986, Merritt 1987, Dejonghe \& Merritt 1988, Meza \& Zamorano
1997) included a variety of models radially anisotropic on the
periphery and isotropic in the center, and vice versa. In these
works, the stability boundaries in terms of the global anisotropy
parameter fall in the broad range $1.2 < \xi < 2.9$. Thus, the
hypothesis about universal stabilization in the narrow region of
$\xi$ was denied. Note, however, that in each case a certain value
of the stability boundary corresponding to the radially
anisotropic system ($\xi > 1$) was found. The same is referred to
the {\it generalized polytropes}, with the DF
\begin{align}
  F(E,L) = C(s,q)\,L^{-s}(-E)^q\quad,
 \label{eq:1.1}
\end{align}
which become stable at $\xi \approx 1.4$ (Fridman \& Polyachenko 1984; Barnes
et al. 1986). Here $E$ and $L$ denote the energy and the angular momentum of a
star, $C(s,q)$ is the normalizing constant, $s$ and $q$ -- parameters of the
model.  Additive constant in gravitational potential $\Phi_0(r)$ is chosen in
such a way that $\Phi_0(R)=0$, where $R$ is  the radius of the system.

Generalized polytropes are the simplest generalization of the isotropic
polytrope models (the latter correspond to $s=0$). The polytrope models
are classical ones in stability theory of both gaseous and collisionless
gravitating systems. One can recall, for example, the work by Antonov (1962),
in which stability of polytrope models with decreasing  DF was shown. Models
with increasing DF can be unstable, but they give exotic mass distribution with
increasing density outwards, thus describing unrealistic stellar systems.

The generalized polytropes are more versatile. The global
anisotropy parameter for (\ref{eq:1.1}) can be obtained in a simple
form:
\begin{align}
 \xi = \frac{2}{2-s}\quad.
 \label{eq:1.2}
\end{align}
Note that for this model the (local) anisotropy parameter
$\beta(r)$ (Binney 1980) does not depend on radius, $\beta =
1-1/\xi = s/2$. It is possible to evaluate expressions for the
radial and the transverse kinetic energy when $s<2$, the limit $s
\to 2$ corresponding to the system in which almost all orbits are
radial. Since $s=0$ case is stable (let us consider only realistic
models with $q>0$), and the case $s\to 2$ is unstable due to the
radial orbit instability, there should be a critical value of the
parameter $s = s_{\rm crit}$ which divides stable and unstable
systems.

Using the matrix method for spheres (Polyachenko \& Shukhman
1981), which is analogous to the Kalnajs matrix method for disks
(Kalnajs 1977), it was found that growth rates of the instability
became small for $s \lesssim 0.6$, almost independently of parameter
$q$ (Fridman \& Polyachenko 1984). Thus, the critical parameter
for generalized polytropes is $s_{\rm crit} \approx 0.6$ (or $\xi
\approx 1.4$). Similar result was obtained by N-body simulations
(Barnes et al. 1986).

Palmer \& Papaloizou (1987) (henceforth PP87) have investigated the same models
using approximate equation for unstable modes with low growth rates.
They showed that instability must persist even for models arbitrary
close to isotropic limit $s=0$; this seemingly contradicts
previous results mentioned above.

Solutions of the approximate equation form a set of infinite number of
unstable modes with decreasing growth rates (eigenvalues) which
accumulate near zero frequency $\omega=0$. These eigenvalues correspond
to eigenfunctions with different number of nodes; the  nodeless
eigenfunction gives the largest growth rate. As was argued in PP87, the
largest eigenvalue cannot be caught by the approximate equation, since
it is too large to fulfill the assumed condition.

Guided by their models B and C, which correspond to $s=2/3$ and
$s=1/3$ ($q=1$), we have calculated the growth rates near the
isotropic limit using the approximate equation. The result was
paradoxical: the largest eigenvalue kept on grow to infinity as $s
\to 0$, while one expects that all modes, including the nodeless
one, would cease to zero.

This paper pursues two goals. First, we try to reconcile
the results obtained by the matrix method and N-body from one side, and
the results of PP87 from the other side. Second, we clarify the paradox
about the applicability of the approximate equation by PP87. For
simplicity, we shall assume models with $s<1$ only. This condition
provides sufficiently smooth gravitational potential in the center, and
linear dependence of the precession velocity on angular momentum for
nearly radial orbits (for more details, see below). In Section 2, we
derive an approximate equation for modes with low growth rates from the
full integral equations for spheres obtained by Polyachenko et al.
(2007). It coincides in all but one important detail with the
approximate integro-differential equation by PP87 (the equivalence
of two equations is demonstrated in Appendix). Numerical results are
given in Section 3, Section 4 contains conclusions.

\section{The integral equation for modes with low growth rates}

Traditional linear stability theories employ matrix methods,
expanding the perturbed potential and density in series using
special biorthonormal sets of basis functions (Kalnajs 1977,
Polyachenko \& Shukhman 1981). As a result, one obtains a set of
integral equations which incorporates the mode frequency in a
complicated nonlinear manner. Thus each frequency is to be
obtained separately by, for example, the Cauchy integration in the
complex plane.

Recently we have proposed an alternative method for calculation of
eigenmodes (Polyachenko 2004, 2005; Polyachenko et al. 2007). The
advantages of our method are (i) linear form of the equation for
eigenmodes and (ii) absence of the basic biorthonormal set that
should be customized for a particular problem. The alternative
method is the most adequate to derive the approximate integral
equation similar to one used in PP87. We start with the full
integral equation for perturbations proportional to spherical
harmonic with the index $l$:
\begin{multline}
\phi_{\,l_1,\,l_2}(E,L) =\frac{4\pi
G}{2l+1}
\sum\limits_{l_1'=-\infty}^{\infty}\sum\limits_{l_2'=-l}^l D_l^{l_2'}
\int\int\frac{dE'\, L'\,dL'}{\Omega_1(E',L')}\,\times \\
\times \,\Pi_{l_1,\,l_2;\,l_1',\,l_2'}(E,L;E',L')
\,\frac{\phi_{\,l_1'\,l_2'}(E',L') {\cal D}_{l'_1,l'_2}F(E',L') }{\omega-\Omega_{l_1'l_2'}(E',L')}.
 \label{eq:2.1}
\end{multline}
Integration in (\ref{eq:2.1}) is over the curved triangle in the
phase plane $(E',L')$: $\Phi_0(0)<E'<0,\ \  0\le L'\le L_{\rm
circ}(E')$; $L_{\rm
circ}(E)$ is the angular momentum on the circular orbit with
energy $E$; $\omega$ is the eigenfrequency; $\Omega_{l_1l_2}(E,L)
\equiv \,l_1\,\Omega_1(E,L)+l_2\,\Omega_2(E,L)$; $\Omega_{1,2}$
are the orbital frequencies; ${\cal D}_{l'_1,l'_2}F(E',L') \equiv
\Omega_{l_1'l_2'}(E',L')\,({\p F}/{\p E'})+l_2'\,({\p F}/{\p
L'})$; the coefficients $D_l^k$ are equal to zero for odd $|l-k|$,
otherwise
$$
 D_l^k=
  \dfrac{1}{2^{2\,l}}
  \,\dfrac{(l+k)!(l-k)!}{\Bigl[\bigl(\frac{1}{2}\,(l-k)\bigr)!\,
 \bigl(\frac{1}{2}\,(l+k)\bigr)!{\phantom{\big|}}\Bigr]^2};
$$
$l_1$ and $l_2$ are indices of  expansion over angular variables $w_1$
and $w_2$  in the action\,--\,angle formalism (Landau \& Lifshitz
1976),
$$ \delta\Phi(I_1,I_2,I_3,w_1,w_2)=\sum\limits_{l_1
l_2}(\delta\Phi)_{l_1l_2}({\bf I})\,\exp[\,i(l_1w_1+l_2w_2)],
$$
conjugated to the action variables $I_i$ ($i=1,2,3$):
$$
I_1=\frac{1}{\pi}\int\limits_{r_{\rm min}}^{r_{\rm
max}}\sqrt{2E-2\Phi_0(r)-\frac{L^2}{r^2}}\,dr,\ \ \ I_2=L,\ \ I_3=L_z.
$$
Due to degeneracy on the azimuthal number $m$, the eigenfrequency
$\omega$ can be calculated for axially symmetric perturbations
$\delta \Phi(r, \theta;t) = \chi(r) P_l(\theta) e^{-i\omega t}$
only. The kernel of the integral equation is
\begin{multline}
\Pi_{l_1,\,l_2;\,l_1',\,l_2'}(E,L;E',L')=\oint dw_1 \oint dw_1'
 \,{\cal F}_l\left[\,r\,(E,L;w_1),r'(E',L';w_1')\right]\times \\
\times
  \cos\Theta_{l_1l_2}(E,L,w_1)\,\cos\Theta_{l_1'l_2'}(E',L',w'_1),
\end{multline}
where $ \Theta_{l_1\,l_2}(E,L;w_1)=({\Omega_{l_1l_2}}/{\Omega_1})\,w_1
-l_2\delta\varphi(E,L;w_1),
$
$$
  \delta\varphi(E,L,w_1) =L\int\limits_{r_{\rm min}(E,\,L)}^{r(E,L,w_1)}
  \frac{dx}{x\,\sqrt{\phantom{\big|}[2E-2\Phi_0(x)]\,x^2-L^2}};
$$
$${\cal F}_l(r,r')={r_<^l}/{r_>^{l+1}}, \quad r_<\equiv{\rm min}
(r,r'), \quad r_>\equiv{\rm max}(r,r').$$
Finally, the eigenfunctions $\phi_{l_1\,l_2}(E,L)$
are connected to the radial part of the perturbed potential as follows:
$$
\phi_{l_1\,l_2}(E,L)=\frac{1}{\pi}\int\limits_0^{\pi}\cos\Theta_{l_1
l_2}(E,L;w_1)\, \chi\bigl[r(E,L,w_1)\bigr]\,dw_1.
$$

\bigskip
To obtain the approximate equation by PP87 from (\ref{eq:2.1}) (rather its
full equivalent in the form of the integral equation in $E$-space), one should
make two simplifications, considering ({\it i}) low frequencies $\omega =
i\gamma$ and even spherical numbers $l$; ({\it ii}) domination of nearly radial
orbits.

 The denominators of resonance terms $l_1'=-\frac{1}{2}\,l'_2$ contain construction
proportional to the precession rate
\begin{align}
 \Omega_{l_1' l_2'}(E',L')=l_2'(\Omega_2-{\case{1}{2}}\,\Omega_1)\equiv l_2'\Omega_{\rm
 pr}(E',L'),
 \label{eq:2.2}
\end{align}
which is small for nearly radial orbits,\footnote{For generalized
polytropes, the gravitational potential behaves like $\Phi_0(r)
\propto r^{2-s}$ near the center. Thus, for the case of our
interest $s<1$ the gravitational force $-\Phi_0'(r)$ is
non-singular at the center and hence the precession velocity of
nearly radial orbits is indeed linear with respect to the angular
momentum, $\Omega_{\rm pr}(E,L) \approx \varpi(E)\,L$ (see, e.g.,
Touma \& Tremaine 1997).  Note that for singular $\Phi_0'(r)$,
(say, $\Phi_0(r)\propto r^{p}$, with $p<1$) the dependence of
precession rate on $L$ is not linear, $\Omega_{\rm pr}\propto
L^p$.}
$$
\Omega_{\rm pr}(E,L)=\varpi(E)\,L.
$$

 Dropping the nonresonance terms and denoting $\phi_{-\frac{1}{2}\,l_2,\,
l_2}(E,0)=\Phi(E)$, from (\ref{eq:2.1}) one can have
\begin{align}
\Phi(E) =-\frac{8\pi G}{2l+1}\sum\limits_{l_2'=2}^l D_l^{l_2'}l_2'^2
\int\int\frac{dE'\, L'\,dL'}{\nu(E')}\,\Phi(E') \,\Pi(E;E') \,\frac{\Omega_{\rm pr}(E',L')\,\dfrac{\p F}{\p
L'}}{\gamma^2+l_2'^2\Omega_{\rm pr}^2(E',L')}.
 \label{eq:2.3}
\end{align}
For DF in the form $F(E,L)=g(E)\,L^{-s}$, the approximate equation
reads
\begin{align}
\Phi(E) =\frac{8\pi G}{2l+1}\sum\limits_{k=2}^l D_l^{k}k^2
\int\int\frac{dE'}{\nu(E')}\,\Phi(E')\,  g(E')\,\varpi(E')\,\Pi(E;E')
\,\int\limits_0 \frac{s\,L^{-s+1}dL}{\gamma^2+k^2\varpi^2(E')\,L^2},
 \label{eq:2.4}
\end{align}
where $\nu(E) \equiv \Omega_1(E,0)$, $\Pi(E,E')$ is the result of
reduction of $\Pi_{l_1,\,l_2;\,l_1',\,l_2'}(E,L;E',L')$ for the
radial orbits.

Due to singularity of DF, the integral in (\ref{eq:2.4}) diverge
when $\gamma=0$, and is large when $\gamma$ is small. This
justifies omission of nonresonance terms, and sets constrains on
the maximum value of  $\gamma$.

It is clear that main contribution to the integral comes from
a narrow region $L \sim \gamma/\varpi$, thus one can change
the variable of integration and replace the upper boundary by infinity:
\begin{align}
{\displaystyle\int_0}
\dfrac{L^{-s+1}dL}{\gamma^2+k^2\varpi^2\,L^2}= \gamma^{-s}
(k\varpi)^{s-2}\,I(s),
 \label{eq:2.5}
\end{align}
where
\begin{align}
I(s)\equiv \int\limits_0^{\infty}
\frac{x^{-s+1}\,dx}{1+x^2}=\frac{\pi}{2\,\sin(\frac{1}{2}\pi\,s)}.
 \label{eq:2.6}
\end{align}

By appropriate change of the eigenfunction, one can reduce the
problem to the integral equation
\begin{align}
 \lambda(s)\,\Psi(E)=\int\limits_{-1}^0 dE'\,{\cal
 R}_s(E,E')\,\Psi(E'),
   \label{eq:2.7}
\end{align}
with
\begin{align}
\lambda=\gamma^s,
 \label{eq:2.8}
\end{align}
and the positively defined symmetric kernel function
\begin{align}
{\cal R}_s(E,E')=\alpha(s)
   \,\sqrt{h(E)\,h(E')\phantom{\big|}}\,Q(E,E'),
     \label{eq:2.10}
\end{align}
where $h(E) = g_s(E) \varpi^{s-1}(E)\, \nu(E)$,
\begin{align}
    Q(E,E')=\int\limits_0^{r_{\rm max}(E)}
  \int\limits_0^{r_{\rm max}(E')}
  \frac{  dr\, dr' {\cal F}_l(r,r')}
  {\sqrt{2E'-2\Phi_0(r')\phantom{\big|}}\,\sqrt{2E-2\Phi_0(r)\phantom{\big|}}},
    \label{eq:2.11}
\end{align}
\begin{align}
\alpha(s)=\frac{(4\pi)^2 G }{(2l+1)}\,\frac{s}{\sin(\pi
s/2)}\,\sum\limits_{k=2}^{l}
 D_l^{k}k^s.
  \label{eq:2.12}
\end{align}
In Appendix, we show that Eg. (\ref{eq:2.7}) in $E$-space is fully
equivalent to the approximate integral equation in $r$-space
obtained by PP87.

The kernel (\ref{eq:2.10}) defines a self-adjoint Hilbert-Schmidt
operator in the infinite-dimensional space, so the eigenvalues
$\lambda_n$ ($n=0,1,2...$) must have an accumulation point,
$\lim\limits_{n\to\infty}\lambda_n=0$. Existence of arbitrary
small eigenvalues is crucial for PP87 in demonstrating the
instability of singular generalized polytropes with $s>0$.

\bigskip
Note that in the limit $s \ll 1$ Eq. (\ref{eq:2.7}) with the
kernel (\ref{eq:2.10}) looks unnatural. Let us consider explicitly
the case $s=0$. The integral equation and the kernel then read as
\begin{align}
   \Lambda\Psi(E)=\int\limits_{-1}^0 dE' {\cal R}_0(E,E')\Psi(E'),
   \label{eq:2.13}
\end{align}
\begin{align}
{\cal R}_0(E,E')=\alpha(0)
\,\sqrt{\frac{g_0(E)\,g_0(E')\,\nu(E)\,\nu(E')}{\varpi(E)\,\varpi(E')}\,\phantom{\big|}}\,Q(E,E'),
\label{eq:2.14}
\end{align}
with  $\Lambda\equiv \lambda(0)$, $\alpha(0)=[32\pi
G/(2l+1)]\,\,\sum\limits_{k=2}^{l} D_l^{k}$. For example, in the
units where $4\pi G=1$, for $l=2$ one has $\alpha(0) = 3/5$. A
norm of the kernel is of order unity and thus first several
eigenvalues, corresponding to eigenfunctions with few nodes, must
be of order unity.
 It is needed to emphasize that there is no small parameter
 left in the problem (\ref{eq:2.13}), the only small parameter
 $s$ in the isotropic limit has disappeared from the equations.

\begin{table}
\begin{center}
 \begin{tabular}{|c|c|c|c|c|c|c|c|}
 \hline
 $l \, \backslash n$ &0&1&2&3&4&5\\
 \hline
    2& {\bf 3.7851} & {\bf 1.0301} &0.4762&0.2637&   0.1621  & 0.1068 \\
   \hline
    4& {\bf 1.3838}  &  0.4194  &  0.2215  &   0.1370  &  0.0921  &  0.0654  \\
   \hline
    6&  0.7029 &  0.2209   &  0.1232  &  0.0803   &  0.0566   &  0.0419  \\
   \hline
 \end{tabular}
\end{center}
\vspace{-2mm}
 \caption{\protect\footnotesize The largest 6 eigenvalues $\Lambda_n$ (i.e. $\lambda_n$ at $s=0$)
 for quadrupole $l=2$, and next two spherical harmonics ($l=4$ and $l=6$)
 for parameter $q=1.0$.
 }
 \end{table}

We have calculated several largest eigenvalues $\Lambda_n$ for
spherical  indices  $l=2,4,6$. The results are summarized in the Table.
The eigenvalues exceeding 1 are emphasized by boldface. In particular,
for $l=2$ two eigenvalues are greater than 1. This means that an
arbitrary small anisotropy (or arbitrary small $s$) will produce
exponentially high growth rates:
\begin{align}
\gamma_n=\Lambda_n^{{1}/{s}}\propto
\exp\Bigl(\frac{1}{s}\,\ln \Lambda_n\Bigr),\ \ \  n=0,1.
  \label{eq:2.15}
\end{align}
However, the growth rates of other modes are exponentially small:
\begin{align}
\gamma_n(s)=\Lambda_n^{\,{1}/{s}}\propto
\exp\Bigl(-\frac{1}{s}\,\ln\frac{1}{\Lambda_n}\Bigr),\ \ n \ge 2.
  \label{eq:2.16}
\end{align}

Note that (\ref{eq:2.15}) contradicts Eq. (\ref{eq:2.3}), in which the kernel
becomes zero at $s=0$ due to the term $\p F/\p L$. The inconsistency evidently
comes from changing the upper limit of integration in (\ref{eq:2.5}) to
infinity: for $s=0$ this integral turns into $\int_0^{L_{\rm max}}{L
dL}/(\gamma^2+k^2\varpi^2\,L^2)$ and diverges if $L_{\rm max} \to \infty$.
However, such a form of the integrand is valid for nearly radial orbits only.
Besides, we have expanded the integration region up to infinite angular
momentum. These are justified for the systems mainly populated by nearly radial
orbits, but not for the nearly isotropic ones.

To cope with anomalously growing modes (\ref{eq:2.15}), one can
take into account the finite value of $L_{\rm max}$ in
(\ref{eq:2.5}). Changing the variable of integration, $L =
[\gamma/(k\varpi)] x$, the integral (\ref{eq:2.5}) can be reduced
to
$$
\int\limits_0^{L_{\rm circ}}
\frac{L^{-s+1}dL}{\gamma^2+k^2\varpi^2\,L^2}=\gamma^{-s}
(k\varpi)^{s-2}\int\limits_0^{{k\varpi L_{\rm circ}}/{\gamma}}
\frac{x^{-s+1}\,dx}{1+x^2}.
$$
Since $\varpi L_{\rm circ} \sim \Omega_2(E,L_{\rm circ}(E))
-\frac{1}{2}\,\Omega_1(E,L_{\rm circ}(E))
=\Omega-\frac{1}{2}\varkappa$, where $\Omega$ and $\varkappa$ are
the circular and radial frequencies, one can replace the upper
boundary of integration by $\Omega/\gamma$, with some
characteristic dynamical frequency $\Omega \sim \Omega_D$,
$\Omega_D \equiv (GM/(2R^3))^{1/2}$. Then instead of
(\ref{eq:2.7}), we have
\begin{align}
\bar I(s, \gamma/\Omega) = \int\limits_0^{\Omega/\gamma} \frac{x^{1-s}dx}{1+x^2} = \frac 12
\int\limits_{\gamma^2/\Omega^2}^1 z^{s/2-1} (1-z)^{-s/2} dz
\label{eq:2:17}
\end{align}
(parameter $\gamma/\Omega \ll 1$). At small $s$, this integral is finite,
$$
\bar I(s, \gamma/\Omega) = s^{-1} [1-(\gamma/\Omega)^s] = s^{-1}[1-\exp(-s\ln(\Omega/\gamma))],
$$
so $\bar I(0, \gamma/\Omega) = \ln (\Omega/\gamma)$, and the kernel turns
to zero at $s=0$. Unfortunately, it is impossible to take the
integral explicitly for arbitrary $s$. Instead, in our crude
approximation, we shall use a model expression obeying necessary features:
\begin{align}
{\bar
I}\bigr(s,{\Omega}/{\gamma}\bigl) = \frac{\pi}{2\sin(\frac{1}{2}\,\pi s)}\,
\Bigl[1-\Bigl(\frac{\gamma}{\Omega}\Bigr)^s\Bigr] = \frac{\pi}{2\sin(\frac{1}{2}\,\pi
s)}\,\Bigl[1-\exp\Bigl(-s\,\ln \frac{\Omega}{\gamma}\Bigr)\Bigr].
 \label{eq:2.17a}
\end{align}
For very small $s$, $s\ll \bigl[\,\ln (\Omega/\gamma)\bigr]^{-1} \ll 1$, the expression gives
$\bar I(s, \gamma/\Omega) \approx \ln (\Omega/\gamma)$, for
$s \gg \bigl[\ln (\Omega/\gamma)\bigr]$ it coincides with old expression (\ref{eq:2.6}).

The corrected approximate integral equation can be written
in the same form as the old one, (\ref{eq:2.7}), (\ref{eq:2.10}),
but instead of (\ref{eq:2.8}) one should write a new relation between $\lambda$ and $\gamma$:
\begin{align}
 \lambda=\frac{\gamma^s}
 {1-\bigl({\gamma}/{\Omega}\bigr)^s},\ \ \
 \gamma=\Bigl(\frac{\lambda}{1+\lambda/\Omega^s}\Bigr)^{\frac{1}{s}}.
  \label{eq:2.18}
\end{align}
This relation clearly provides exponentially small growth rates
for all modes at $s\to 0$. In this limit, the spectrum of
eigenvalues $\lambda_n(s)$ remains the same, but now
\begin{align}
 \gamma_n(s)\propto \left(\frac{\Lambda_n}{1+\Lambda_n}\right)^{1/s}=
 \exp\Bigl[-\frac{1}{s}\,\ln\Bigl(1+\dfrac{1}{\Lambda_n}\Bigr)\Bigr]\
   \label{eq:2.19}
\end{align}
asymptotically tends to zero when $s\to 0$.

\section{Unstable modes of generalized polytropes}

For numerical evaluations of growth rates of anisotropic  polytropes,
one usually introduces the following units:
\begin{align}
   4\pi G=1,\ \   A(s,q) = 1,\  \Psi(0)= 1,
   \label{eq:3.1}
\end{align}
where $G$ is the gravitational constant, $\Psi(r)$ is the relative potential,
$\Psi(r) = -\Phi_0(r)$, and $A(s,q)$ is the coefficient in the expression for density distribution,
\begin{align}
   \rho_0(r) = A(s,q)\,r^{-s}\,\Psi^{\,q+\frac{3-s}{2}}.
   \label{eq:3.2}
\end{align}
The relation between $A(s,q)$ and $C(s,q)$ is well-known (see, e.g., Fridman \&
Polyachenko 1984):
\[
A(s,q)=C(s,q)\,(2\pi)^{3/2}\,2^{-s/2}\,\frac{\Gamma(q+1)\,\Gamma\bigl(-\frac{1}{2}\,s+1\bigr)}
{\Gamma\bigl(q+\frac{1}{2}(5-s)\bigr)},
 \]
$\Gamma(x)$ denotes the Gamma function.

In these units, the dimensionless potential $\psi(r) = -\Phi_0(r)$ satisfies the Poisson equation:
\begin{align}
   \frac{d^2\psi}{dr^2}+\frac{2}{r}\,\frac{d\psi}{dr}+r^{-s}\,\psi^{\,q+\frac{3-s}{2}}=0,
   \label{eq:3.3}
\end{align}
with a boundary condition $\psi(0)=1$. To find a second boundary
condition, one can notice that the Poisson equation possesses a
solution in terms of a power series in $z = r^{2-s} \equiv r^p$  (see
also H\'{e}non 1973). Denoting $\bar \psi (z) = \psi(r)$, one has
\begin{align}
\frac{d^2{\bar\psi}}{dz^2}+\frac{3-s}{2-s}\,\frac{1}{z}\,\frac{d{\bar\psi}}{dz}+\frac{1}{(2-s)^2}\,\frac{1}{z}\,
{\bar\psi}^{\,q+\frac{3-s}{2}}=0.
 \label{eq:3.4}
\end{align}
In the limit $z \ll 1$, the solution is ${\bar\psi}(z)=1+D_1
z+{\cal O}(z^2)$, where $D_1=-1/[(2-s)(3-s)]$. Thus, one can
obtain the potential by integrating (\ref{eq:3.4}) from $z=0$,
with the boundary conditions $\bar\psi(0) = 1$,
$\left({d{\bar\psi}}/{dz}\right)\Big|_{z=0} = D_1$. The right
boundary of integration $z_R$ is determined from the condition
$\bar \psi (z_R) = 0$, thus radius of the system is $R =
z_R^{1/(2-s)}$.

Evaluation of the radial frequency $\nu(E) \equiv \Omega_1(E,0)$
is usual. The precession rate at low angular momenta ($\varpi(E) =
(\p\Omega_{\rm pr}(E,L)/\p L)|_{L=0}$) is calculated according to
Eq. (2.11) by Polyachenko et al. (2007):
$$ \varpi(E)=\frac{\nu(E)}{\pi}\Biggl[\int\limits_0^{r_{\rm
max}(E)}\frac{dr}{r^2}\Bigl(\frac{1}{\sqrt{2E+2\Psi(r)\phantom{\big|}}}
-\frac{1}{\sqrt{2E+2\phantom{\big|}}}\Bigr)- \frac{1}{r_{\rm
max}(E)\sqrt{2E+2\phantom{\big|}}}\Biggr].
$$
The following asymptotic formulas, applicable for stars in the very
center of the sphere, $\varepsilon \equiv E+1 \ll 1$, are used for
testing of  the precision of calculations:

\medskip\noindent\tbs $s \ll 1$

$$
 \varpi(\varepsilon)={\frac{1}{40}}\,\Bigl(q+{\frac{3}{2}}\Bigr)\,
 \Bigl[1+{\frac{1}{160}}\,\varepsilon\,
\Bigl(1-{\frac{262}{7}}\,q\Bigr)\Bigr]+\frac{s}{12\,\varepsilon};
$$

\medskip\noindent\tbs otherwise, $s < 1$

\begin{align}
\nu(\varepsilon) &=\sqrt{2\pi}\,\,
\frac{\Gamma(\frac{1}{p}+\frac{1}{2})}{\Gamma(1+\frac{1}{p})}\,
K^{\frac{1}{p}}\,\varepsilon^{\frac{1}{2}-\frac{1}{p}}; \\
\varpi(\varepsilon) &=
-\frac{\Gamma(\frac{1}{2}+\frac{1}{p})\,\Gamma(1-\frac{1}{p})}
{\Gamma(1+\frac{1}{p})\,\Gamma(\frac{1}{2}-\frac{1}{p})}\,K^{\frac{2}{p}}\,
\varepsilon^{-\frac{2}{p}}>0,
 \label{eq:3.5}
\end{align}
where $K^{-1}=p\,(1+p)$, $p=2-s$.

Some problems were experienced in calculating the function $Q(E,E')$
(\ref{eq:2.11}), which is a part of the kernel (\ref{eq:2.10}). The
following asymptotic formulas are useful (valid for $l=2$):

\medskip\noindent\tbs on the diagonal $\varepsilon=\varepsilon'\ll 1$, $s=0$
$$Q(\varepsilon,\varepsilon) \simeq \sqrt{{\frac{3}{2}}}({\frac{1}{2}}+{\bf G})
\,\varepsilon^{-1/2}=
1.734\,\varepsilon^{-1/2},$$
where ${\bf G}=0.915\,965\,594...$ is the Catalan's constant;

\medskip\noindent\tbs $\varepsilon'\ll\varepsilon\ll 1$, arbitrary $s$

\begin{align}
Q(\varepsilon,\varepsilon')=\frac{5\,\sqrt{\pi}}{12p}\,
 \frac{\Gamma(\frac{1}{p})}{\Gamma(\frac{1}{p}+\frac{1}{2})}\,K^{-\frac{1}{p}}\,
 \varepsilon^{-\frac{1}{2}}\,
(\varepsilon')^{\frac{1}{p}-\frac{1}{2}}.
 \label{eq:3.6}
\end{align}

The integral equation (\ref{eq:2.7}) have been solved for modes with the
spherical harmonics $l=2,4,6$. First of all, we were interested in models with
small parameter $s \ll 1$ to find out how small  the growth rates of
weakly anisotropic systems are. Also we have calculated growth rates for
models B ($s=2/3$, $q=1$) and C ($s=1/3$, $q=1$) of PP87. In the numerical
work, we restricted ourselves by models within the range $0 \le s \le 0.8$.

\begin{figure}
\begin{center}
 \includegraphics[width=87mm, draft=false]{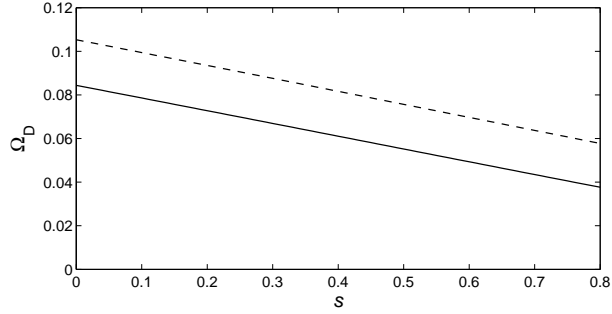}
 \end{center}
\vspace{-4mm}
 \caption{The dependence of characteristic frequency
 $\Omega_D$ on the parameter $s$ for $q=1$ (solid line) and $q=0.7$ (dashed line).}
 \label{fig1}
\end{figure}

Fig.\,\ref{fig1} shows the behavior of the characteristic
frequency $\Omega_D(s) = (GM/2R^3)^{1/2}$ for $q=1$ and $q=0.7$.
One can see that the dependence is weak and $\Omega_D \sim 0.1$ in
the considered range of the parameter $s$.

\begin{figure}
\begin{center}
 \includegraphics[width=87mm, draft=false, bb = -18 197 614 543, clip=]{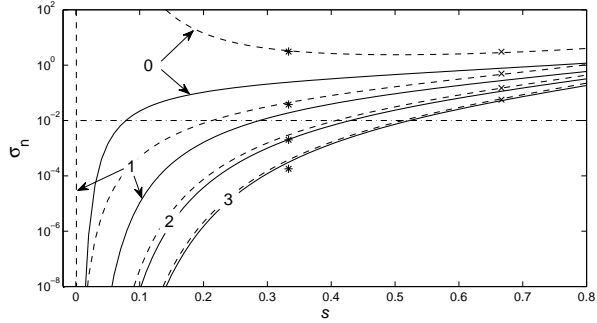}
 \end{center}
\vspace{-4mm}
 \caption{The dependence of the scaled growth rates
 $\sigma_n \equiv \gamma_n/\Omega_D$ of four most unstable modes on the
 parameter $s$ for the model $q=1$: dashed curves show solutions
 for (\ref{eq:2.8}), solid curves -- for the corrected relation (\ref{eq:2.18}).
 Crosses show values of $\sigma_n$ for model B, stars -- for model C from PP87.
 The characteristic frequency is $\Omega = 0.08$.}
 \label{fig2}
\end{figure}

Fig.\,\ref{fig2} shows the growth rates $\sigma_n$ in units of
characteristic frequency $\Omega_D$ v.s. the parameter $s$ for
$q=1$. Dashed lines show growth rates obtained from (\ref{eq:2.7})
with relation (\ref{eq:2.8}); this case is equivalent to the
approximate equation used by PP87. For $s=1/3$ and $s=2/3$, our
growth rates agree satisfactorily with one obtained in the cited
paper. Numbers denote modes (0 -- the nodeless mode, 1 -- the mode
with one node, etc.). The growth rates of the first two modes
increase violently as the model approach the isotropic limit,
other modes decrease exponentially to zero.

Solid lines show the growth rates v.s. the parameter $s$ obtained from
(\ref{eq:2.18}). Its behavior complies with intuitive expectations
that the unstable modes should be stabilized in the isotropic
limit.

\begin{figure}
\begin{center}
 \includegraphics[width=87mm, draft=false, bb = -18 197 614 543, clip=]{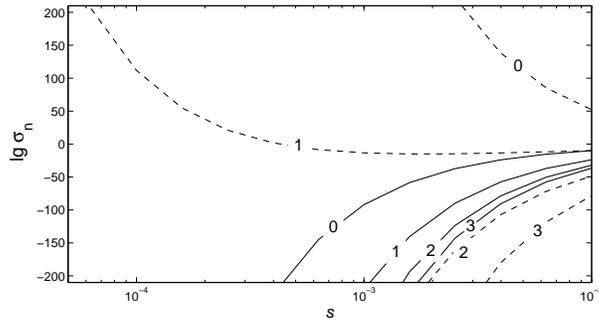}
 \end{center}
\vspace{-4mm}
 \caption{The same as in Fig.\,\ref{fig2}, in lg -- lg axis. }
 \label{fig3}
\end{figure}

A region near $s \approx 0$ is magnified in Fig.\,\ref{fig3}. Due
to fast (exponential) decrease of the growth rates, they become
negligibly small at $s = 0.01$.

\begin{figure}
\begin{center}
 \includegraphics[width=87mm, draft=false, bb = -18 197 614 543, clip=]{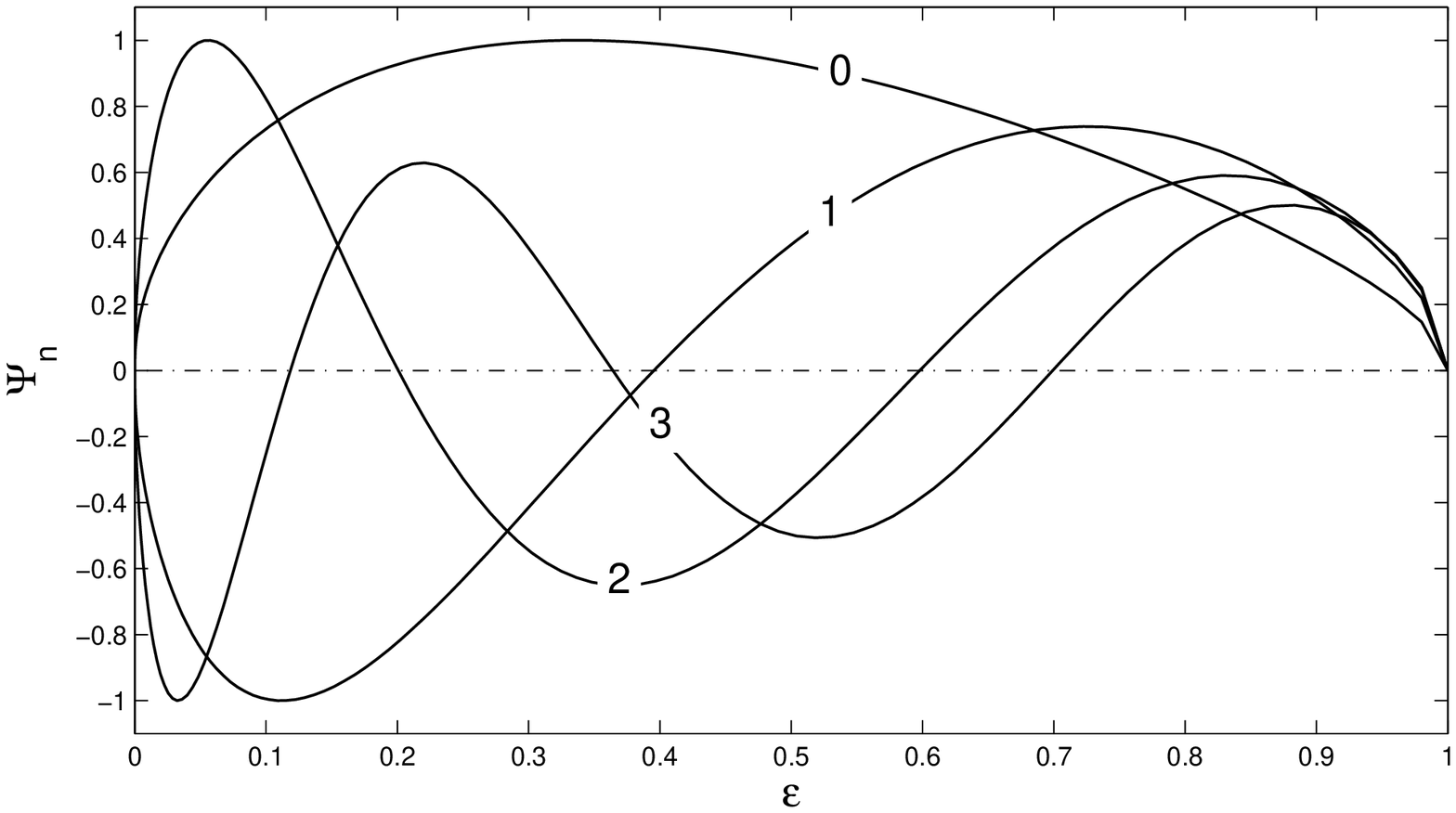}
 \includegraphics[width=87mm, draft=false, bb = -18 197 614 543, clip=]{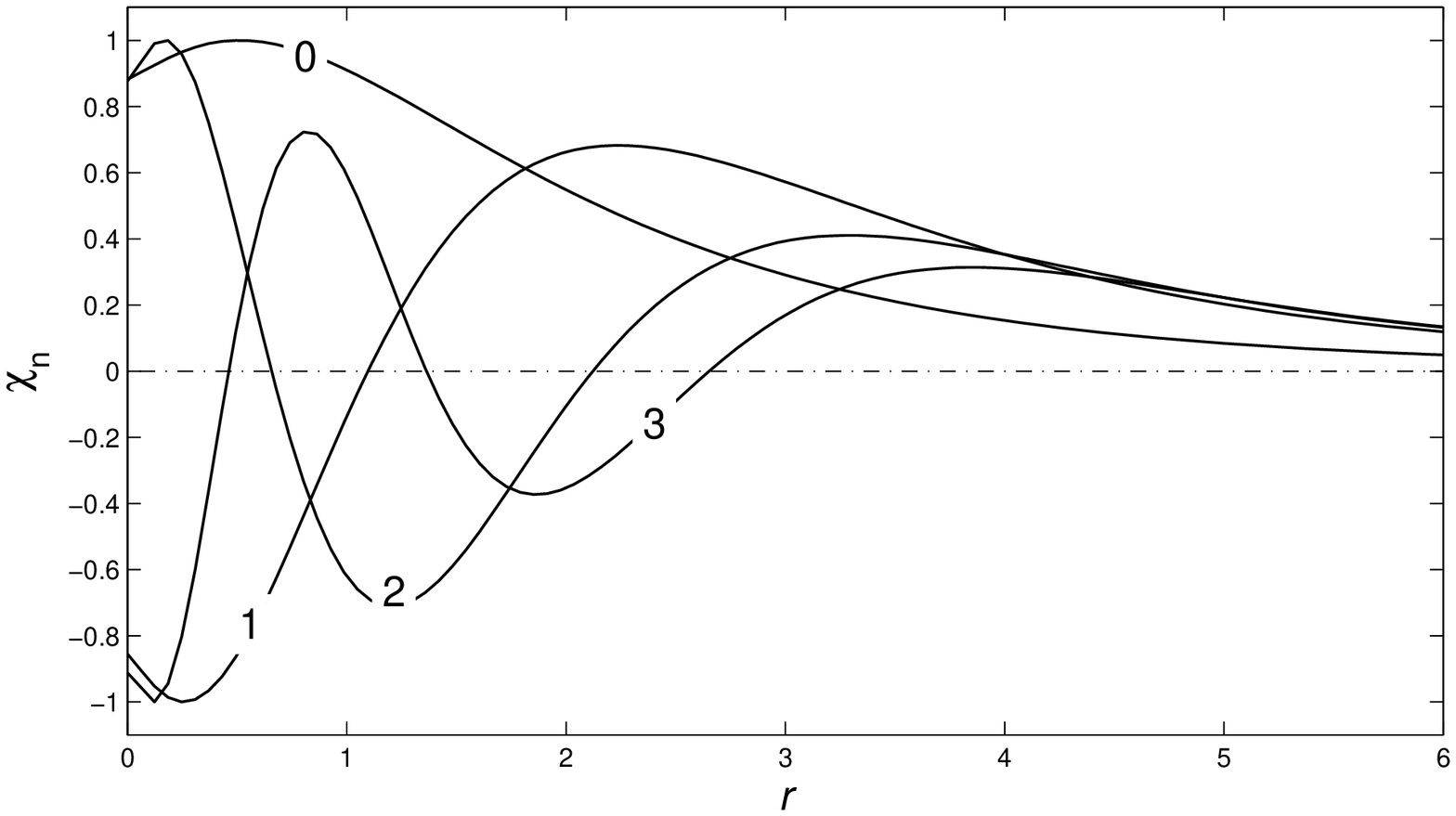}
 \end{center}
\vspace{-4mm}
 \caption{ The first four eigenfunctions of quadrupole harmonics $l=2$:
 $\Psi_n(E)$ (upper figure) and the radial part of the potential $\chi_n(r)$
 (lower figure). The normalization of the eigenfunction is arbitrary.}
 \label{fig4}
\end{figure}

Fig.\,\ref{fig4} shows several first eigenfunctions $\Psi_n(E)$ ($n=0,1,2,3$)
of the integral equation (\ref{eq:2.7}) for the model $s=1/3$, $q=1$
and corresponding radial parts of the potential $\chi_n(r)$:
\begin{align}
\chi_n(r) = \int\limits_{-1}^0 \d E \,
 \sqrt{h(E)}\, \Psi_n(E)\!\!\!\int\limits_0^{r_{\rm max}(E)}\!\!\!\!
 \frac{dr'\,\,{\cal F}_l(r,r')}{\sqrt{2E-2\Phi_0(r')\phantom{\big|}}}
   \label{eq:3.7}
\end{align}
(for derivation of this relation, see Appendix). In all cases, the
eigenfunctions have equal number of nodes coinciding with $n$. The form of
radial parts of the potential $\chi_n(r)$ is in qualitative agreement with
those presented in Fig. 1 of PP87, which are the solutions of the integral
equation (2.25).

\begin{figure}
\begin{center}
 \includegraphics[width=87mm, draft=false, bb = -18 197 614 543, clip=]{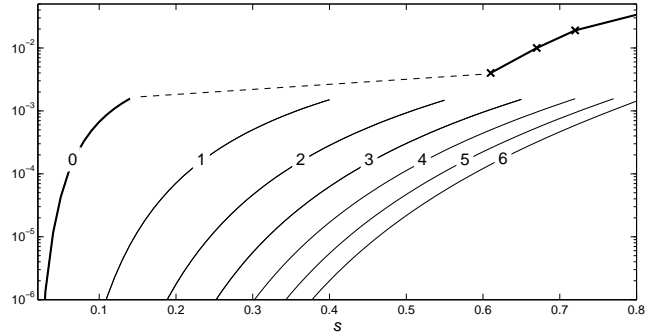}
 \end{center}
\vspace{-4mm}
 \caption{The dependence of the growth rates $\gamma_n$ from (\ref{eq:2.18})
 of the first seven unstable modes on parameter $s$ for the model $q=0.7$.
 Crosses mark the growth rates obtained by Fridman \& Polyachenko (1984).
 The characteristic frequency is taken $\Omega = 0.02$. }
 \label{fig5}
\end{figure}

It is interesting to compare the growth rates obtained with our approximate
integral equation with independent calculations of generalized polytropes. In
Fig.\,\ref{fig5}, we show the dependence of growth rates for first seven modes
on the parameter $s$ for $q=0.7$. Crosses mark an ``experimental'' curve from
Fridman \& Polyachenko (1984). The curve breaks at $s \approx 0.6$, $\gamma
\approx 0.004$ because of the accuracy of the matrix method employed. The
numbers  poorly agree with each other, so that it is impossible to join the curve
marked by crosses with a new curve of the principal mode $n=0$.

The validity of the approximate integral equation is restricted to
very small growth rates. For modes with sufficiently high $n$,
this restriction does not play any role since their growth rates
are small for all $s$, but it is important for modes with few
nodes, especially for the nodeless one. Its growth rate increases
with anisotropy, and presumably at $s \lesssim 0.15$ achieves the
upper boundary. This might be the reason of the evident
discrepancy between the two curves. In Fig.\,\ref{fig5}, the
uncertainty region $0.15 \lesssim s < 0.6$ is shown by a dashed
line.

\section{Conclusions}

1. It is difficult to reconcile the results of the linear
stability analysis (Fridman \& Polyachenko, 1984) and N-body
experiments (Barnes et al. 1986) with the results by PP87 for two
reasons.

First, the approximate integral equation derived in PP87, on which their
analysis is based, is applicable to very small growth rates only. In the
isotropic limit corresponding $s \to 0$, these growth rates are exponentially
small, i.e. $\gamma \propto \exp(-s_{\ast}/s)$.  The estimates show, that even
for $s \approx 0.5$, the allowed growth rates are much less than $0.01
\Omega_D$, that is definitely below any reasonable accuracy of the matrix
method and N-body experiments.

Second, strictly speaking, the approximate integral equation is not
applicable to unstable modes having eigenfunctions with just few nodes,
including the principal mode with maximum growth rate,  since for not
too large $n$ the eigenvalues $\lambda_n$ are not very small even in
the limit $s\to 0$. However the principal mode  plays the major role in
determining the stability boundary (see Fridman \& Polyachenko (1984),
Barnes et al. (1986)).

2. The growth rates formally accurate when they are small. However, it is likely that
once $\sigma_n\equiv \gamma_n/\Omega_D < 1$, they give reasonable estimates of
actual growth rates in practice (PP87). Then since isotropic systems with
decreasing DF are stable (Antonov 1960, 1962), {\it all} modes  should become
stable when $s$ approaching zero, and valid approximate equation must describe
{\it all} modes correctly. This fact is in contradiction to our solution  of
PP87's approximate equation (2.25) (equivalent of our Eq. (\ref{eq:2.7}) in
which the relation $\lambda=\gamma^s$ is assumed). This solution demonstrates
explicitly that for the quadrupole harmonic $l=2$ there are two modes with
exponentially increasing growth rates for $s\to 0$.

The reason for such a discrepancy arises from behavior of the
terms of the approximate integral equation in the limit $s \to 0$.
Using the notation of PP87, the growth rates can be expressed in
the form
$$
\gamma^s = \frac{A_3}{A_1 - A_2},
$$
where $A_i$ are some averages (quadratic forms) of the positively
defined operators.\footnote {For convenience, we have changed
signs of all three PP87's operators ${\cal L}_i$ to make them
positively defined.}
 In particular, $A_3$ denotes the average of the integral operator defined by
(\ref{eq:2.10}),  $A_2$ denotes the omitted nonresonance part and
$A_1$ denotes the operator which is a left side of the radial Poisson
equation. Since both terms $A_1$ and $A_3$ retain in the limit
when the system becomes isotropic, Palmer \& Papaloizou infer that
instability exists no matter how weak the divergence in DF as $L
\to 0$ is.

In this paper we argue that the term $A_3$ {\it must vanish} when $s\to
0$ in order to comply with stabilization of isotropic models. Using our
alternative method of determining the unstable eigenmodes based on
solution of the linear eigenvalue problem, we derived the appropriate
integral equation. This equation gives a set of unstable modes, all of
which become stable in the isotropic limit $s \to 0$.

\bigskip

\noindent  To summarize, our considerations prove that the instability
growth rates of all modes in unbounded models at $L=0$ indeed do not
vanish unless the models are isotropic, but they becomes exponentially
small. Actually it means stability if we take into account a finite
lifetime of real astronomical objects. Besides, the most probable
distributions are non-singular ones, so we have to infer that stable
distributions generally become unstable at some {\it finite} value of
radial anisotropy, i.e. finite anisotropy threshold exists. Width of
the threshold depends on a particular model.

\section*{Acknowledgments}

The work was supported in part by Russian Science Support Foundation, RFBR
grants No. 11-02-01248, No.~09-02-00082, No.~10-05-00094 and also by Programs
of Presidium of Russian Academy of Sciences No 16 and OFN RAS No. 16.

\subsection*{\large APPENDIX A. Equivalence of the integral equation in $E$-space and the integro-differential equation
(2.25) of PP87 in $r$-space}
\medskip

Actually, the integro-differential equation (2.25) of PP87
presents the radial part of the Poisson equation for $l$-th
harmonic of the potential
$$
\frac{1}{r^2}\,\frac{d}{dr}\,r^2\,\frac{d\chi(r)}{dr}-\frac{l(l+1)}{r^2}\,\chi(r)=4\pi\,
G\,\Pi(r),
 \eqno{(A1)}
$$
where $\Pi(r)$ is the radial part of a density perturbation
$\delta\rho(r,\theta;t)=\Pi(r)P_l(\cos\theta)\,e^{-i\omega t}$. We
can calculate it using an expression for the perturbed DF
$\delta\!f$ obtained from the kinetic equation written in the
actions\,--\,angles variables (Landau \& Lifshitz 1976). For
details of calculation of $\delta\!f$ see  Polyachenko \& Shukhman
(1981) or Fridman \& Polyachenko (1984). Thus we have
$$
\Pi(r)P_l(\cos\theta)=\int d\bv \,\delta\!f(\br,\bv)=$$
$$=-\sum\limits_{l_1}\sum\limits_{l_2} \int d\bv
P_l^{l_2}(0)P_l^{-l_2}(\sin\theta_0)e^{il_2\pi}\phi_{l_1l_2}(E,L)
 \frac{l_1\,{\p F}/{\p I_1}+l_2\,{\p F}/{\p
I_2}}{\omega-\Omega_{l_1l_2}}\,e^{i(l_1w_1+l_2w_2)},
 \eqno{(A2)}
$$
where $\sin\theta_0=L_z/L$.  Separation of the radial part of
perturbed density yields:
$$
\Pi(r)={\case{1}{2}}\,(2l+1)\int\limits_0^{\pi}
\delta\rho(r,\theta)P_l(\cos\theta)\,\sin\theta\,d\theta.
 \eqno{(A3)}
$$
We can  integrate over $\theta$ using addition theorem for Legendre polynomials
and a relation which connects $\theta$  with the angular variable $w_2$
$$
\cos\theta=\cos\theta_0\cos(w_2-\p S_1/\p I_2),
 \eqno{(A4)}
$$
where $S_1$ is radial action,
 $$
 S_1(r)=\int_{r_{\rm min}}^r dr'\,
\sqrt{2E-2\Phi_0(r')-L^2/r'^2}.
$$
 We have
$$
P_l(\cos\theta)=\sum\limits_{k=-l}^{l} e^{-ik(w_2-\p S_1/\p
I_2)}P_l^k(\sin\theta_0)\,P_l^{-k}(0)\, e^{-ik\pi}.
 \eqno{(A5)}
$$
 Since
$$
v_r=\pm\,\sqrt{2E-2\Phi_0(r)-\frac{L^2}{r^2}},\ \ v_{\theta}=\pm
\frac{L}{r}\sqrt{1-\frac{\sin^2\theta_0}{\sin^2\theta}},\ \ v_{\varphi}=
\frac{L}{r}\,\frac{\sin\theta_0}{\sin\theta},
$$
then we obtain for the volume element in the velocity space
$$
d\bv\equiv dv_r dv_{\theta}dv_{\varphi}=\frac{4L\,dL\,dE\,
d(\sin\theta_0)}{r^2\,\sqrt{2E-2\Phi_0(r)-{L^2}/{r^2}\phantom{\big|}}\,\,
\sqrt{\cos^2\theta_0-\cos^2\theta\phantom{\big|}}}.
 $$
(The factor 4 appears here because we have to take into account
particles having velocities $v_r$ and  $v_{\theta}$ of both
signs.) To integrate over $\theta$ in  (A3) it is convenient to
go to integration over $w_2$. We have
$$
\sin\theta\, d\theta=\cos\theta_0\,\sin(w_2-\p S_1/\p I_2)\,dw_2
$$
and
$$
\int\limits_0^{\pi}\frac{P_l(\cos\theta)\,e^{il_2w_2}\,\sin\theta\,d\theta}
{\sqrt{\cos^2\theta_0-\cos^2\theta\phantom{\big|}}}=
{\case{1}{2}}\int\limits_0^{2\pi}P_l(\cos\theta)\,e^{il_2w_2}dw_2.
 \eqno{(A6)}
$$
As a result we obtain
$$
\Pi(r)=-{\case{1}{4}}\,(2l+1)(2\pi)\sum\limits_{l_1}\sum\limits_{l_2}\int\int
4L\,dL dE \, \frac{\phi_{l_1l_2}(E,L)\,e^{i(l_1w_1+l_2\p S_1/\p
I_2)}} {r^2\,\sqrt{2E-2\Phi_0(r)-{L^2}/{r^2}\phantom{\big|}}} \frac{l_1\,{\p
F}/{\p I_1}+l_2\,{\p F}{\p I_2}}{\omega-\Omega_{l_1l_2}}\times$$
$$\times
\int\limits_{-1}^{1}  dz
P_l^{l_2}(0)\,P_l^{-l_2}(z)\,P_l^{-l_2}(0)\,P_l^{l_2}(z),\ \ z\equiv
\sin\theta_0.
 \eqno{(A7)}
$$
 Taking into account that the integral over  $z$ is
$$
\int\limits_{-1}^{1}dz\,P_l^{l_2}(0)P_l^{-l_2}(z)P_l^{-l_2}(0)P_l^{l_2}(z)=
\frac{2}{2l+1}\,D_l^{l_2},
$$
we find the final expression for the radial part of perturbed
density $\Pi(r)$:
 $$
 \Pi(r)=-\frac{4\pi}{r^2}\sum\limits_{l_1}\sum\limits_{l_2=-l}^{l}
 D_l^{l_2}\int\int
L\,dL\,dE\, \frac{\phi_{l_1l_2}(E,L)\,\cos{\Theta_{l_1l_2}}}
{\sqrt{2E-2\Phi_0(r)-{L^2}/{r^2}\phantom{\big|}}}\,\frac{\Omega_{l_1l_2}\,{\p
F}/{\p E}+l_2\,{\p F}/{\p L}}{\omega-\Omega_{l_1l_2}}\,,
 \eqno{(A8)}
 $$
where (see Sec. 2)
 $$
\phi_{l_1\,l_2}(E,L)=\frac{1}{2\pi}\oint\limits_0^{\pi}\cos\Theta_{l_1
l_2}(E,L;w_1)\, \chi\bigl[r(E,L,w_1)\bigr]\,dw_1=
\frac{\Omega_1}{\pi}\int\limits_{r_{\rm min}}^{r_{\rm max}}\frac{dr\,\chi(r)
 \cos\Theta_{l_1l_2}}{\sqrt{2E-2\Phi_0(r)-{L^2}/{r^2}\phantom{\big|}}}.
   $$
Keeping in the sum over  $l_1$ the resonance summand
$l_1=-\frac{1}{2}\,l_2$ and supposing growth rate to be small, we
retain the contribution with $\p F/\p L$ only:
$$
 \Pi(r)\approx \frac{8\pi}{r^2}\sum\limits_{l_2=2}^{l} D_l^{l_2}l_2^2\int
L\,dL\int\frac{\phi_{-l_2/2,\,l_2}(E,L)\,dE}
{\sqrt{2E-2\Phi_0(r)-{L^2}/{r^2}\phantom{\big|}}}\,\,  \frac{\Omega_{\rm pr}\,{\p F}/{\p L}}{\gamma^2+l_2^2\Omega_{\rm
pr}^2}\,\cos{\Theta_{-l_2/2,\,l_2}}.
 \eqno{(A9)}
 $$
Since the leading contribution to the integral over $L$ comes from small $L$,
we put where it is possible, $L=0$ and suppose $\Omega_{\rm pr}=\varpi(E)\,L$.
Then $\Theta_{-l_2/2, l_2}(E,0)=l_2\pi$, and finally
 $$
 \Pi(r)=-\frac{8\pi s\,I(s)}{\gamma^s}\frac{1}{r^2}
 \sum\limits_{l_2=2}^{l} D_l^{l_2}l_2^s
 \int\frac{g(E)\,\varpi^{s-1}(E)\,dE} {\sqrt{2E-2\Phi_0(r)\phantom{\big|}}} \Biggl[
 \frac{\nu(E)}{\pi}\int\limits_{0}^{r_{\rm max}(E)} \frac{dr' \chi(r')}
{\sqrt{2E-2\Phi_0(r')\phantom{\big|}}}\Biggr].
 \eqno{(A10)}
 $$
Substitution to r.h.s. of (A1) yields integro-differential equation (2.25) of
PP87 in $r$-space:
 $$
\frac{d}{dr}\,r^2\,\frac{d\chi(r)}{dr}-l(l+1)\,\chi(r) =
-\frac{4\pi\,G}{\gamma^s}\,\int\limits_0^R K(r,r')\,\chi(r')\,dr',
 \eqno{(A11)}
$$
with the kernel
$$
K(r,r')=\frac{4\pi s}{\sin(\pi\,s/2)}
 \sum\limits_{k=2}^{l} D_l^{k}k^s 
 \int\limits_{{\rm max}\bigl[\Phi_0(r),\Phi_0(r')\bigr]}\!\!\!\!
 \frac{g(E)\,\varpi^{s-1}(E)\,\nu(E)\,dE} {\sqrt{2E-2\Phi_0(r)\phantom{\big|}}\,
  \sqrt{2E-2\Phi_0(r')\phantom{\big|}}}.
   \eqno{(A12)}
$$
Now it is easy to demonstrate that (A11) is equivalent to the
integral equation (2.8) in $E$-space. We write (A10) in the form
$$
 \Pi(r)=-\frac{4\pi^2 s}{\sin(\pi s/2)}\frac{1}{\gamma^s}\frac{1}{r^2}
 \sum\limits_{k=2}^{l} D_l^{k}k^s\!\!
 \int\frac{g(E)\,\varpi^{s-1}(E)\,dE} {\sqrt{2E-2\Phi_0(r)\phantom{\big|}}}\,\Phi(E),
 \eqno{(A13)}
 $$
where
 $$
 \Phi(E)=
 \frac{\nu(E)}{\pi}\int\limits_{0}^{r_{\rm max}(E)} \frac{dr' \chi(r')}
{\sqrt{2E-2\Phi_0(r')\phantom{\big|}}}
 \eqno{(A14)}
 $$
is the radial part of perturbed potential  $\chi(r)$ averaged over
radial orbit. Rewriting the Poisson equation (A1) in the integral
form,
  $$
  \chi(r)=-\frac{4\pi G}{2l+1}\int dr' r'^2\,\Pi(r')\,{\cal
  F}_l(r,r'),
    \eqno{(A15)}
  $$
and averaging  both part of this equation over radial orbit with
the energy $E$ according to (A14), we obtain the integral equation
 $$
 \Phi(E)=\frac{1}{\gamma^s}\int dE' {\cal K}(E,E')\,\Phi(E'),
   \eqno{(A16)}
 $$
where
 $$
  {\cal K}(E,E')=\frac{(4\pi)^2 G }{(2l+1)}\,\frac{s}{\sin(\pi s/2)}
  \,g(E')\,\varpi^{s-1}(E')\,\nu(E)\times$$ $$\times
  \sum\limits_{k=2}^{l} D_l^{k}k^s\!\!\!\!
  \int\limits_0^{r_{\rm max}(E)}\!\! dr\!\!\!\!
  \int\limits_0^{r_{\rm max}(E')}\!\!\!\!\!\!\! dr'\,
  \frac{{\cal F}_l(r,r')}
  {\sqrt{2E'-2\Phi_0(r')\phantom{\big|}}\,\sqrt{2E-2\Phi_0(r)\phantom{\big|}}}\,.
   $$
Symmetrizing this equation with the help of the substitution
   $$
\Psi(E)=\sqrt{\frac{g(E)\,\varpi^{s-1}(E)}{\nu(E)}}\,\Phi(E),
$$
we obtain the integral equation (2.8) with the kernel (2.10)
presented in the main text.
\bigskip

Finally, using the expression (A13) for $\Pi(r)$ and the integral
form of the Poisson equation (A15), we can easily obtain the
relation (\ref{eq:3.7}), which connects the eigenfunction
$\Psi_n(E)$ of the integral equation (\ref{eq:2.7}) with the
eigenfunction $\chi_n(r)$ of PP87's integro-differential equation
(2.25) (or Eq. (A11) in our notations).

\end{document}